\documentclass[aps, prx, 10pt, amssymb,amsmath,superscriptaddress,tightenlines,twocolumn,notitlepage] {revtex4-2}
\usepackage{hyperref}
\usepackage{tabularx}
\usepackage{xurl}
\usepackage{physics}
\usepackage{amsmath}
\usepackage[dvipsnames]{xcolor}
\usepackage{amsfonts}
\usepackage{placeins}

\usepackage{geometry}
\usepackage{graphicx} 

\begin{document}
%\title{Berry Curvature Measurement via Ballistic Transport}
\title{Angle-Resolved Berry Curvature via Nonlinear Hall Effect of Ballistic Electrons}
\author{Louis Primeau}
\affiliation{Department of Physics and Astronomy, University of Tennessee, Knoxville, TN, USA}

\author{Qiong Ma}
\affiliation{Department of Physics, Boston College, Chestnut Hill, MA, USA}

\author{Yang Zhang}
\affiliation{Department of Physics and Astronomy, University of Tennessee, Knoxville, TN, USA}
\affiliation{Min H. Kao Department of Electrical Engineering and Computer Science, University of Tennessee, Knoxville, Tennessee, USA}

%\date{\today}
%\affiliation{%
% Department of Physics, University of Tennessee
%}

\begin{abstract}
Berry curvature fundamentally dictates the topological ground state, anomalous transport and optical properties of quantum materials. However, directly mapping its momentum-space distribution in real materials remains an outstanding experimental challenge. Here, we present an inverse method for reconstructing the abelian Berry curvature of a single band using angle-resolved measurements of the transverse conductance. Our inversion relies on a symmetry-constrained statistical model with two hyperparameters that can be inferred directly from the nonlinear Hall conductance, yielding a parameter-free inversion method. We demonstrate the feasibility of our method using simulated measurements of tight-binding models of WSe$_2$ and $ABC$-stacked trilayer graphene. 
\end{abstract}

\maketitle

\textit{Introduction.} Electrons in a periodic crystal are described by Bloch wavefunctions, for which not only the energy dispersion (energy–momentum eigenvalues) but also the underlying geometric properties—such as Berry curvature and the quantum metric—play essential roles~\cite{Girvin_Yang_2019}. These geometric quantities, together with the band structure, provide key ingredients for understanding and predicting instabilities toward ordered phases, including superconductivity and magnetism.

Experimentally, the characterization of electronic energy dispersion is well established. In particular, angle-resolved photoemission spectroscopy (ARPES)~\cite{damascelli2003angle,hufner2013photoelectron,sobota2021angle} provides direct and momentum-resolved access to the electronic structure. By photoemitting electrons with incident photons, ARPES measures the energy and momentum of the emitted electrons, thereby directly mapping the occupied band structure in reciprocal space~\cite{damascelli_arpes, nano8050284, Yang2018, Zhang2020ARPES}. Extending such momentum-resolved capability to quantum geometric quantities would provide a powerful means to interrogate electronic structure in quantum materials. Recent work has made progress along this direction, demonstrating access to the quantum geometric tensor via polarization-resolved photoemission, although limitations remain for directly resolving the momentum-dependent details~\cite{kang2025measurements,kim2025direct}.

Transport measurements, particularly Hall responses, provide an alternative route to probing quantum geometric properties, as they are intrinsically sensitive to Berry curvature. The linear anomalous Hall effect probes the momentum-integrated Berry curvature~\cite{nagaosa2010anomalous}, while the second-order nonlinear Hall effect is sensitive to the Berry curvature dipole~\cite{sodemann2015quantum,ma2019observation,Du2021,SHEN2024100535}, corresponding to the first moment of Berry curvature over the Fermi surface. Higher-order nonlinear Hall responses can, in principle, access higher moments, such as the Berry curvature quadrupole~\cite{sankar2024experimental,ma2021topology}. However, these approaches inherently involve momentum averaging over occupied states or the Fermi surface.

Here, we propose an alternative approach based on the ballistic regime of nonlinear Hall transport. This approach enables momentum-selective probing of Berry curvature and is naturally compatible with nanodevice architectures, thereby providing a pathway to overcome the intrinsic momentum averaging of diffusive transport measurements and to achieve momentum-resolved access to quantum geometric quantities.

\begin{figure}
    \centering
    \includegraphics[width=1.0\linewidth]{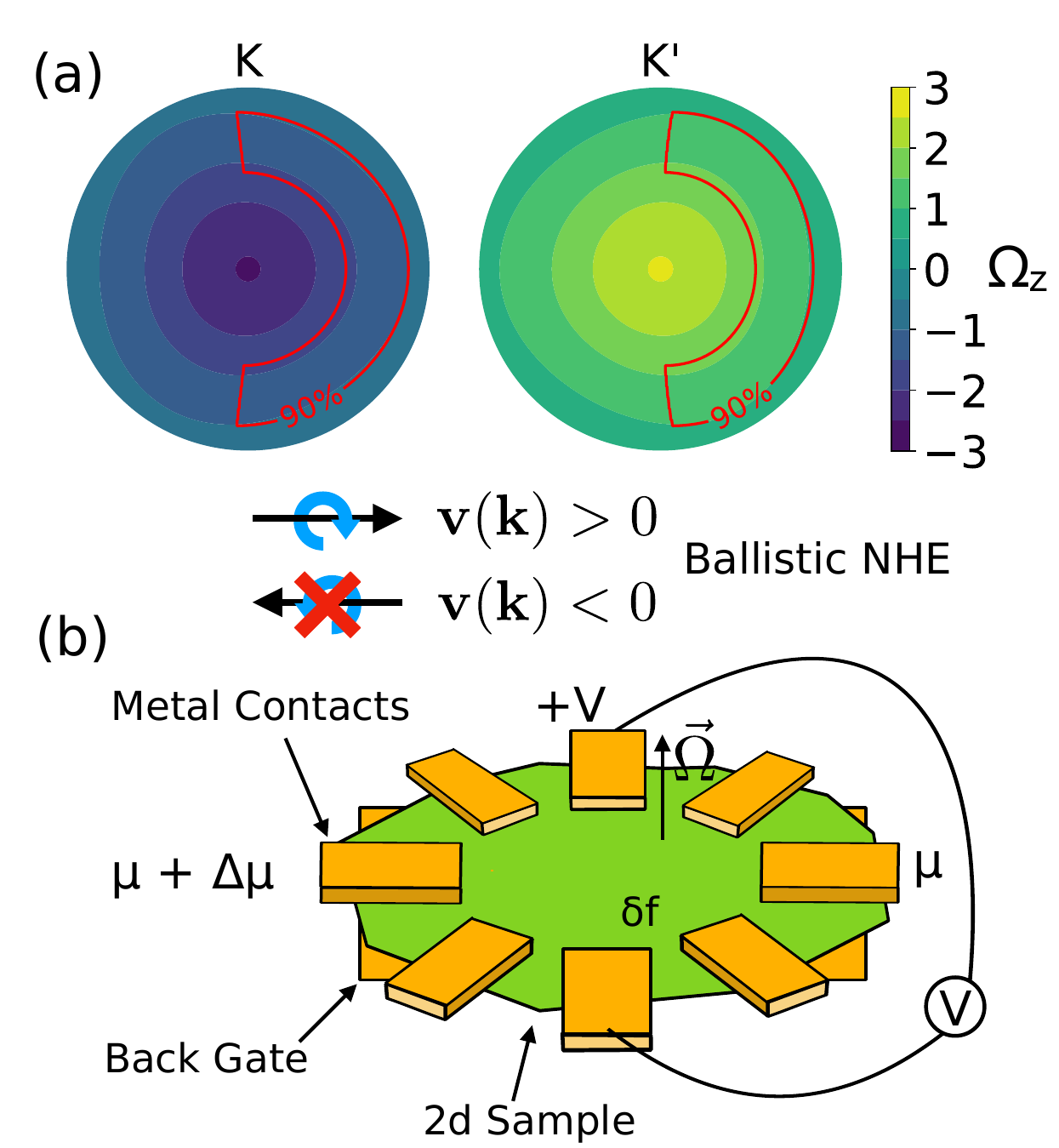}
    \caption{Part (a) shows the Berry curvature at the $K$ and $K'$ valleys of WSe$_2$. The surplus distribution is at the Fermi surface with positive velocity. Electrons with below the Fermi surface or with negative velocity do not contribute to transport. The 90\% contours of the distribution $\delta f = -\frac{\partial n_F}{\partial \varepsilon}\Theta(v_x)$ are shown in red ($k_B T = 0.02$ eV). In our inversion we consider chemical potentials in which the Fermi surface of WSe$_2$ is entirely within these two regions. Part (b) shows the proposed measurement setup. For each angle we have a pair of probes which inject and accept electrons. Orthogonal to these are floating probes which measure the Hall current. The injecting probes are switched to resolve different angles.}
    \label{fig:setup}
\end{figure}

\begin{figure*}[t]
    \centering
    \includegraphics[width=0.8\linewidth]{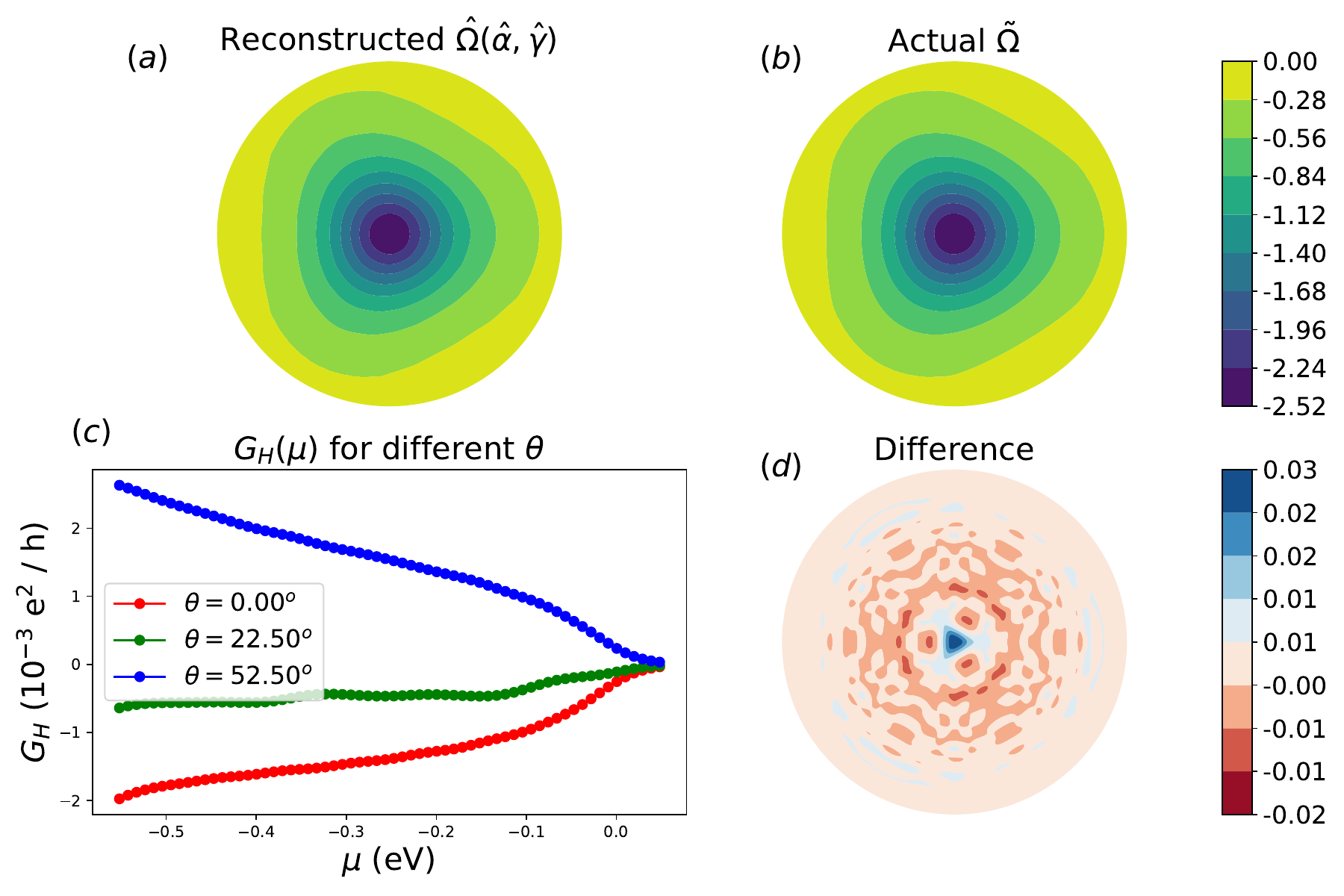}
    \caption{Reconstruction of the Berry curvature (BC) in the $K$ pocket of WSe$_2$. In (a) is the reconstructed Berry curvature, and (b) is the true Berry curvature calculated from the tight-binding model. In (c) is the difference between these two fields. Points not constrained by the data have been masked out of the plot. The hyperparameter tuning chose the regularizing parameter $\hat \gamma = 0.0034$. The region of largest error is close to the $K$ point. A detailed analysis of the dependence of the reconstruction and the spectrum on the regularizing parameters is provided in the appendix.  }
    \label{fig:wse2}
\end{figure*}

In clean materials with a long mean free path, low-temperature ballistic transport measurements are known to reveal directional information about band structure and topology~\cite{magnushalleffect,PhysRevLett.128.076801, PhysRevB.57.8907, PhysRevB.99.035440}. Specifically, the nonlinear ballistic Hall effect~\cite{SHEN2024100535} can selectively probe the Berry curvature at the Fermi surface~\cite{magnushalleffect}. In a ballistic setup of length $L$ sandwiched between two reservoirs (metallic leads at fixed chemical potential), the anomalous transverse current is given by
\begin{equation}
I^{yxx}_{H} = E_x^2\frac{e^3 L^2}{h 2\pi} \int d^2 k \left(-\frac{\partial n_F}{\partial \varepsilon}\Theta(-\mathbf{v}(\mathbf{k}) \cdot \mathbf{E})\right) \Omega_z(\mathbf{k})
\label{eq:continuous_operator}
\end{equation}
where $\Omega_z$ is the $z$-Berry curvature, $\mathbf{v}(\mathbf{k})$ is the group velocity, $n_F$ is the Fermi-Dirac function centered at the chemical potential $\mu$. We designate $\mathbf{E}$ to be at an angle $\theta$ relative to the $x$-axis of the lab coordinate system. The key term $\Theta(-\mathbf{v}(\mathbf{k}) \cdot \mathbf{E})$ is the Heaviside step function which enforces the fact that in a ballistic setup excess electrons flow only in the direction opposite to the electric field. By comparing conductance measurements while varying the chemical potential and electric field direction, the Berry curvature can be determined everywhere in the Brillouin zone.

An intuitive argument for why Equation \ref{eq:continuous_operator} can be used to determine the Berry curvature is as follows. First we define the effective linear response $G_H = I^{yxx}_H / eE_x L$ and chemical potential imbalance $\Delta \mu=eE_xL$ which drives the current. If we fix the chemical potential and perform an ideal measurement at zero temperature, what we measure is the integral of the Berry curvature along one segment of the Fermi surface contour determined by $\mu$ and $\theta$. If we perform a second measurement of $G_H$ with a small offset in $\theta$, then subtracting these two measurements yields the difference in Berry curvature between the beginning and the end of the contour. By collecting measurements at many angles, the Berry curvature along the whole Fermi surface can be determined. To move the contour to a different energy we can shift the Fermi surface by modifying $\mu$. Of course, knowledge of the shape and location of the contour, which is given by the band structure, must be determined beforehand either by the experimental methods described above or first-principles calculations. The possibility of reconstruction depends on the details of the Fermi surface shape. For example, if we have a pair of time-reversal (TR) related circular Fermi surfaces, then the integral along the contour on each will perfectly cancel out and we get zero contribution from the ballistic Hall response. In this case the Berry curvature is totally unconstrained. Fortunately, in real materials with anisotropic Fermi surfaces, the contribution from the two TR related pockets do not cancel at all angles and the Berry curvature can be reconstructed.

At finite temperature, the smearing of the Fermi function relates Berry curvatures at neighboring energies to the conductance measurement at a single chemical potential $\mu$. Figure \ref{fig:setup}a shows the region in the Brillouin zone involved in the Hall conductance measurement for a given $\mu$ and $\theta$. In the following we propose an experimental setup and derive a regularized least-squares method to combine measurements of $G_H$ at different angles and chemical potentials. The method is designed for finite temperature. We then demonstrate it on tight-binding models of WSe$_2$~\cite{PhysRevB.88.085433} and $ABC$-stacked rhombohedral graphene~\cite{zhang2010band,chen2019signatures,zhou2021superconductivity,yang2022spectroscopy} when the measurements are corrupted by noise.

\textit{Experimental Setup and Statistical Inverse Model.} The proposed experimental setup is sketched in Figure \ref{fig:setup}b. The basic idea is to exploit the step function term $\Theta(-\mathbf{v}(\mathbf{k}) \cdot \mathbf{E})$ in order to gain angular resolution on the Berry curvature. The sample of clean 2D material is laid on top of a ring of contacts patterned onto a substrate. A pair of opposite contacts is used as the source and drain in a 4-probe measurement of the transverse conductance $G_{yx}$. The difference in chemical potential between the source and drain determines the magnitude $\Delta \mu$ of the nonlinear Hall conductance, and the lower of the two sets the overall chemical potential $\mu$. Two orthogonal probes serve as floating voltage probes. The number of contacts determines the angular resolution of the measurements. Finally, a back gate is used to apply a $z$-electric field to break inversion symmetry if it is needed, as in the case of rhombohedral graphene. The size of the sample must be large enough that the semiclassical approximation is valid, yet smaller than the mean free path so that the electrons travel through the sample without scattering. 

\begin{figure*}
    \centering
    \includegraphics[width=0.85\linewidth]{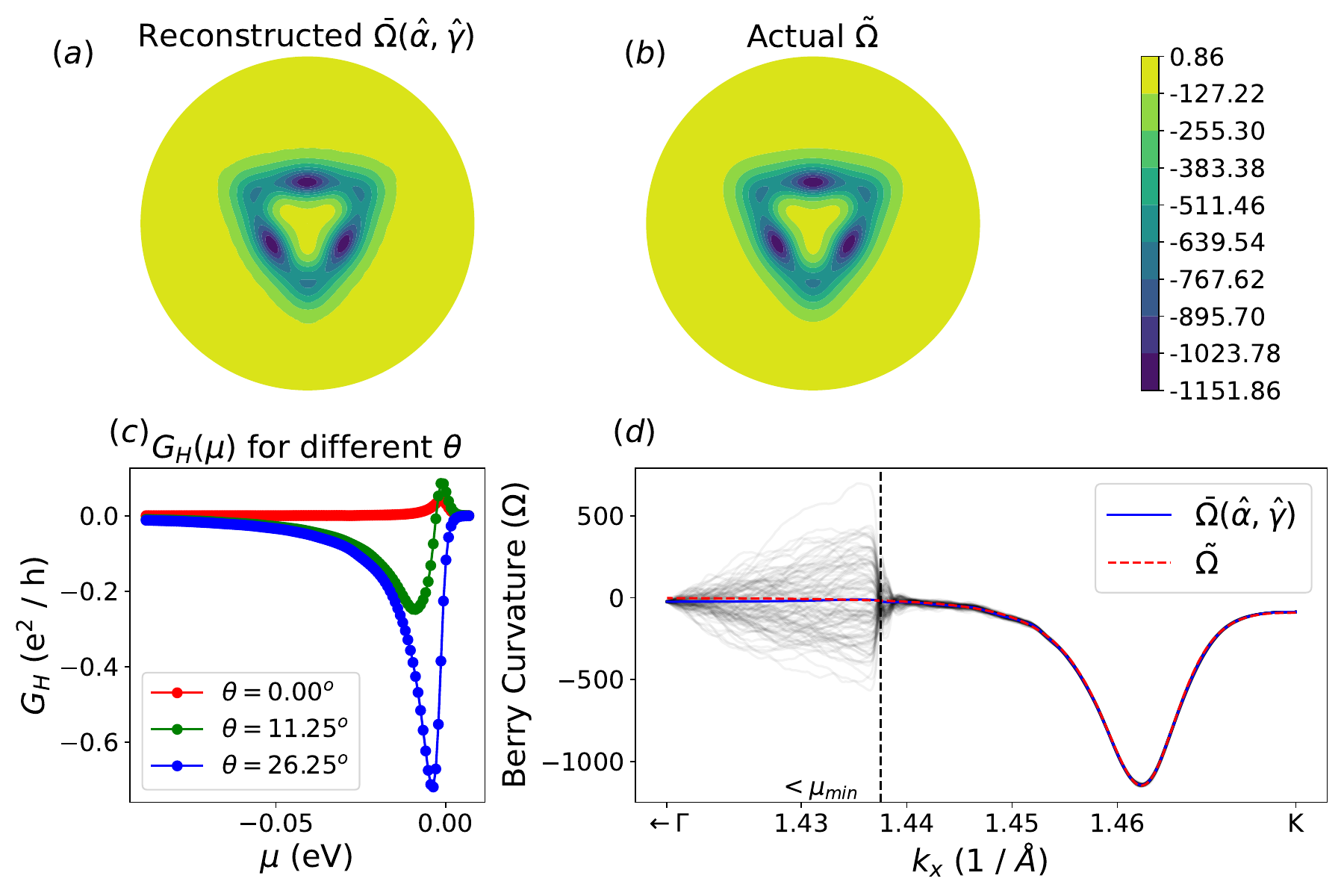}
    \caption{Reconstruction of the Berry curvature with low noise in the $K$ pocket of ABC-stacked graphene. The $K'$ pocket is determined by time-reversal symmetry, which is explicitly enforced in the inversion procedure. In (a) is the reconstructed Berry curvature and (b) is the true Berry curvature from the tight-binding model. (c) shows the difference between these two fields. The grid search of the log posterior that led to the hyperparameters is shown in the supplementary materials. In (c) examples of the Hall response are shown at different angles used for the inversion. Part (d) shows the MAP estimate and true Berry curvature for the low-noise case of ABC-stacked graphene. Random draws from the marginal posterior are shown to estimate the error of the measurement. For draws to the right of the dashed vertical line, the $k$-point has energy within the chemical potentials measured. The variance is much larger for points outside this range, shown to the left of the dashed line. }
    \label{fig:abc}
\end{figure*}
As a first step to performing the inversion, other contributions to the conductance due to Fermi surface anisotropy must be removed from the measurement of $G_{yx}$. These contributions are discussed in the supplementary materials (section V), and include terms constant and linear in $\Delta \mu$. To get the linear component we can fit a line through the measured conductance as a function of $\Delta \mu$ for each angle and chemical potential. Once this is done, the non-anomalous contribution can be subtracted off and we are left with a signal described by Equation \ref{eq:continuous_operator}.

Discretization of Equation \ref{eq:continuous_operator} onto meshpoints $\mathbf{k}_j$ leads to the expression
\begin{equation*}
G_H(\mu, \theta) =  a(\mu, \theta, \mathbf{k}_j)^T \Omega(\mathbf{k}_j)
\end{equation*}
where $a(\mu, \theta, \mathbf{k}_j)^T$ is the row vector corresponding to the discretized integral for chemical potential $\mu$ and angle $\theta$ (supplementary materials section VI). The experiment consists of measuring the response at a set of angles and chemical potentials $(\mu, \theta)_i$ by varying the potentials of the various leads. Instead of performing the subtraction procedure between two angles as described above, we simply stack all the row vectors together to form a matrix $A$ and all the measurements together to form the vector $G^o$, so that in the absence of other signals we expect
\begin{equation*}
G^o =  \sum_{j} A((\mu, \theta)_i,  \mathbf{k}_j) \Omega(\mathbf{k}_j)
\end{equation*}
From here on out we drop the $i$ and $j$ subscripts and use matrix notation. We have the statistical model
\begin{equation}
G^o = A \Omega + \epsilon
\label{eq:statistical_model}
\end{equation}
where the measurement process is corrupted by a zero-mean Gaussian random variable $\epsilon$ with covariance $(\alpha W)^{-1}$. Here $W$ describes the correlation structure of the noise in the observations and $\alpha$ is an adjustable scale parameter on the noise level. If we do not have any reason to believe that the errors are correlated, then we choose $W$ as a multiple of the identity, which we do in all numerical experiments. The parameter $\alpha$ allows us to estimate the amount of noise. Without any priors, this leads to the maximum likelihood estimate for $\Omega$. However, without further assumptions the matrix $A^T W A$ is very singular (see supplementary Figure \ref{fig:svd} for the singular value spectrum). One reason is that the mathematical space of functions that satisfy Equation \ref{eq:continuous_operator} is too large because it includes functions which can vary arbitrarily fast. These functions can be ruled out on physical grounds because the Berry curvature is a smooth function away from a band crossing. Another reason is that we have a finite number of measurements. To enforce these physical constraints, we use a Laplacian $L$ on the irreducible wedge which penalizes high frequency variation and removes grid-scale numerical noise. The operator is constructed using finite differences satisfying the exact discrete conservation laws and Neumann boundary conditions at all boundaries. With this prior, the mean of the posterior distribution conditioned on $\alpha$ and $\gamma$ is given by
\begin{equation}
\bar \Omega(\alpha, \gamma) = \Lambda^{-1} A^T \alpha W G^o
\label{eq:maximum}
\end{equation}
where $\Lambda \equiv \alpha A^T W A + \gamma L^T L$ for prior width parameter $\gamma$, which is the regularizing hyperparameter of the statistical model. Here for a fixed $\alpha$, increasing $\gamma$ acts as a scale cutoff. The measurement of symmetry-equivalent angles drastically reduces the error (by $1 / \sqrt{n}$ for $n$ equivalent measurements). This greatly increases the robustness of the inversion to corrupting noise. It is not clear \textit{a priori} how to set the value of the hyperparameter $\gamma$, which in principle depends on the level of noise. It turns out that both $\gamma$ and $\alpha$ can be automatically inferred from the data~\cite{mackay, MacKay1996} by maximizing the marginal log posterior

\begin{widetext}
\begin{align}
\begin{split}
\log P(\alpha, \gamma | G^o) &= \frac{1}{2} \log \det \alpha W - \frac{1}{2}(G^o - A\bar{\Omega})^T \alpha W (G^o - A \bar{\Omega})  \\
&+ \frac{1}{2}\log\det \left(\gamma L^T L \right) - \frac{1}{2}\log \det\Lambda - \frac{1}{2}\bar{\Omega} (\gamma L^T L) \bar{\Omega}.
\label{eq:marginal_log_likelihood}
\end{split}
\end{align}
\end{widetext}
\noindent
This expression is derived in the supplementary materials and involves integrating out $\Omega$ which is possible because the posterior is Gaussian in $\Omega$. The first line is the log of the best-fit likelihood and the second line is the log of the Occam factor~\cite{mackay}. The MAP estimate~\cite{10.1093/oso/9780198568315.001.0001}  $\hat{\alpha}$ and $\hat{\gamma}$ are obtained by maximizing this expression, which are then used to calculate $\hat \Omega = \bar\Omega(\hat \alpha, \hat \gamma)$ using Equation \ref{eq:maximum}. For our synthetic data we know the true solution $\tilde{\Omega}$ so we will be interested in the error
\begin{equation*}
e = |\tilde \Omega - \hat{\Omega}|
\end{equation*}
where the norm $|\cdot|$ will be taken with respect to the area elements of the coordinate system. In particular for our synthetic data we can check the hyperparameter tuning by plotting the true error $e(\gamma, \alpha) = |\tilde\Omega - \bar{\Omega}(\alpha, \gamma)|$ as a function of the hyperparameters and comparing with the marginal log-likelihood. The maximum of the marginal log posterior typically coincides with the minimum curve of the error (it is a curve because the error depends only on the ratio $\gamma / \alpha$, see Figure \ref{fig:log_likelihood}), leading to good reconstruction quality. 

\textit{Numerical Demonstration.} To demonstrate that our inversion method works, we made synthetic data by running the forward model, i.e., multiplying the true Berry curvature calculated in a tight-binding model by $A$, then adding Gaussian noise as described by Equation \ref{eq:statistical_model}. First we will demonstrate that the inversion procedure works in the low-noise limit, which will show the best possible performance of the inverse model. Then we will check the large-noise limit to see at what point the inversion becomes unreliable.

We begin with the TR-symmetric transition metal dichalcogenide WSe$_2$. We used a 3-band model from~\cite{PhysRevB.88.085433} with spin-orbit coupling to calculate the band structure and the true Berry curvature from which we generate the synthetic data. In this material the valence band has maxima at the $K$ and $K'$ points which are related by TR symmetry and therefore have opposite Berry curvature. The material has three-fold rotational, mirror-$x$, and time-reversal symmetries, which we exploit in the inversion procedure as we will describe below. 

The synthetic data consisted of simulated measurements on a linearly spaced Cartesian product of chemical potentials and angles with 64 points between -0.551 eV to 0.049 eV for $\mu$ and 48 points from $0$ to $2\pi$ for $\theta$. The temperature broadening $k_B T$ was set to 20 meV which is on the order of the measurement spacing in $\mu$, however the reconstruction was not sensitive to the exact temperature. The noise level was set to $0.01 \times \frac{1}{n}\sum_{n} |G_H|$. 

With these parameters the Fermi surface is near the $K$ and $K'$ points. To generate the numerical mesh for the construction of $A$ and $L$ we explicitly enforced all the listed symmetries by performing the reconstruction on an irreducible wedge with the tip at the $K$ point. The numerical meshes used for the synthetic data generation and the inversion procedure were kept the same.

The reconstruction is shown in Figure \ref{fig:wse2}. The maximum relative error $|\tilde{\Omega} - \hat{\Omega}| / \tilde{\Omega}$ was below $4\%$ for points fully constrained by the data, i.e., having energy $\varepsilon(\mathbf{k})$ within the bounds of the measurement mesh in $\mu$. For points outside this range, $\hat{\Omega}(\mathbf{k})$ is masked out in the plots. The largest error is at the $K$ point, and decreases radially outward. Overall there is excellent agreement between the reconstructed and true Berry curvature.

Next we consider a model of three-layer $ABC$-stacked graphene. We break the inversion symmetry by applying an out of plane gating field of 0.1 V/\AA, which amounts to adding a $z$-varying onsite term in the tight-binding model. The band maxima are shifted away from the $K$ and $K'$ points so that the Fermi surface has a three-pocket structure. This material also has three-fold rotational, mirror-$x$, and time-reversal symmetries. In this material the band structure and the Berry curvature do not vary together monotonically, and the maximum of the Berry curvature is not at the same location as the band maximum. The chemical potential was varied from 88.1 to 6.9 meV (again 0 eV is the valence band maximum). The thermal broadening was set to 1 meV.  The same number of points for the measurements in $\mu$ and $\theta$ were used as in the previous example. Once again all symmetries were used to generate the irreducible wedge centered at the $K$ point.

The reconstruction is shown in Figure \ref{fig:abc}. The reconstruction is able to recover the three pocket structure accurately with some errors in the details near the $K$ point. For this material the Berry curvature is near zero on some mesh points, and hence the relative error measure is not applicable. We can however state that the range of the error is small compared to the range of Berry curvature on the grid, about 5\%, and refer the reader to the difference plot in Figure \ref{fig:abc}c. In this example a grid search was performed over $\gamma$ and $\alpha$ to maximize the marginal log posterior (Equation \ref{eq:marginal_log_likelihood}). Contours of the log-posterior and the error $e(\alpha, \gamma)$ are shown in the supplementary material (Figure \ref{fig:log_likelihood}). The automatic hyperparameter tuning procedure correctly estimates the noise scale $\alpha$ used to corrupt the data and the smoothing hyperparameter. 

The above two low-noise examples demonstrate that the prior is reasonable and that the inversion procedure is possible. Now we move on to the realistic situation in which the data is corrupted by noise, for example intrinsic noise in the measurement or residual misfit from the subtraction of other signals. Here we keep the same synthetic generation setup as in the previous case of ABC-stacked graphene, but add noise with a standard deviation equal to $0.5 \frac{1}{n}\sum |G_H|$.

The reconstruction for a noisy signal is shown in Figure \ref{fig:noisy_reconstruction}. An example synthetic measurement is shown in Figure \ref{fig:noisy_reconstruction}(a) in which noisy and clean data are plotted together to show the scale of noise tolerable by the inversion. This measurement is taken off the mirror-symmetric angle so it has maximum $G_H$. The reconstruction is able to capture quantitative and qualitative features, including the correct location of the three-fold pocket. It also captures the order of magnitude of the Berry curvature. This example shows that our inversion technique is robust to even severe noise in the signal. 

Finally, the reconstruction is improved by taking more measurements in $\mu$ and $\theta$, as shown in the supplementary materials Figure \ref{fig:error_scaling}. We found that reasonable reconstruction accuracy could be obtained for as few as 12 measurements in $\theta$ and $\mu$.

Our inversion relies heavily on knowledge of the band structure, which we assumed was known perfectly. In reality, the band structure can be modified by many-body effects and may not match first-principles calculations. In future work we could improve the reconstruction procedure by modeling uncertainty in $\varepsilon(k)$ either from imperfect calculations or measurement noise. Another factor is the idealization of the geometry used to derive Equation \ref{eq:continuous_operator}. Realistic modeling of the metal-sample contacts and scattering from the sample boundaries would improve the inversion procedure in a real experiment, but is outside the scope of this work.

\textit{Conclusion.} In this work, we have discussed a novel method for determining the Berry curvature in clean samples of two-dimensional materials. We have numerically demonstrated that the signal from ballistic nonlinear Hall transport~\cite{SHEN2024100535} contains enough information to determine the Berry curvature to good accuracy with only simple assumptions on its smoothness.The parameter-free Bayesian reconstruction successfully extracts intricate topological features---such as the multi-pocket Berry curvature distribution in ABC-stacked graphene---even under severe experimental noise. 

While our current inversion assumes a precisely known energy dispersion $\varepsilon(\mathbf{k})$, in realistic 2D devices, especially at low density, band structures are sometimes modified by many-body renormalizations or substrate-induced strain. Future extensions of this technique could systematically incorporate uncertainty in $\varepsilon(\mathbf{k})$ directly into the Bayesian prior, effectively treating the energy dispersion and quantum geometry as jointly inferred quantities. Furthermore, while our derivation assumes uniform local electric fields, finite contact widths and mesoscopic boundary scattering in realistic device geometries will blur the angular resolution. The built-in Laplacian regularization minimizes these finite-size artifacts by enforcing physical smoothness, ensuring the topological reconstruction remains robust against non-ideal contact configurations.

Ultimately, this angle-resolved transport protocol serves as a blueprint for a momentum-space topological microscope. It opens a pathway to directly image localized quantum geometry in systems where standard spectroscopic probes are severely limited, such as encapsulated van der Waals heterostructures or actively gated moiré superlattices. Tracking the evolution of $\Omega_{z}(\mathbf{k})$ across gate-tuned phase diagrams will provide crucial, momentum-resolved signatures of interaction-driven topological transitions and renormalized band structures in the Fermi-liquid phase.

\FloatBarrier
\bibliography{refs}

\pagebreak
\widetext

%%%%%%%%%% Merge with supplemental materials %%%%%%%%%%
%%%%%%%%%% Prefix a "S" to all equations, figures, tables and reset the counter %%%%%%%%%%
\setcounter{equation}{0}
\setcounter{figure}{0}
\setcounter{table}{0}

\makeatletter
\renewcommand{\theequation}{S\arabic{equation}}
\renewcommand{\thefigure}{S\arabic{figure}}
\renewcommand{\bibnumfmt}[1]{[#1]}
\renewcommand{\citenumfont}[1]{#1}
\setcounter{page}{1}

\begin{center}
\textbf{\large Supplemental Materials}
\end{center}

\section{Additional Terms}
Here we extend the ballistic transport results of \cite{magnushalleffect} to second-order to account for corrupting signals. We also want to explicitly list the assumptions so that the experimental design respects the approximations. We begin by deriving the boundary conditions. Following ~\cite{magnushalleffect} we approximate the geometry as a strip of length $L$ in the $x$ direction and infinite in the $y$ direction. On the left and right boundaries there are two reservoirs at chemical potentials $\mu_L$ and $\mu_R$ that send out electrons at their thermal distributions. The electrons are not scattered in the sample and dephase into the equilibrium distribution of the other source as soon as they reach it. The left movers on the right boundary have two types: those with enough initial energy to overcome the potential barrier and make it to the left reservoir, and those that do not. We assume that the chemical potentials $\mu_L$, $\mu_R$ of the reservoirs are much larger than their difference $\Delta \mu$ so that we can ignore the electrons without sufficient kinetic energy. In the absence of a magnetic field, the electrons obey the equations of motion
\begin{align*}
\dot{\mathbf{r}} = \frac{1}{\hbar} \nabla_{\mathbf{k}} \varepsilon - \frac{e}{\hbar}\boldsymbol{\Omega}(\mathbf{k}) \times \mathbf{E}(\mathbf{r}) \\
\dot{\mathbf{k}} = -\frac{e}{\hbar} \mathbf{E}(\mathbf{r}).
\end{align*}
 For a two-dimensional system in the $x$-$y$ plane the Berry curvature is given by $\boldsymbol{\Omega} = (0, 0, \Omega_z)$, and we assume an electric field induced by the spatially varying electrochemical potential:
 \begin{equation*}
 E(x) = \frac{1}{e}\frac{\partial \mu}{\partial x}.
 \end{equation*}
 The steady-state, collisionless Boltzmann equation in a $y$-translationally invariant system obeying the the above equations of motion is
\begin{equation}
\left( \frac{1}{\hbar} \frac{\partial \varepsilon}{\partial k_x} \frac{\partial}{\partial x} - \frac{1}{\hbar} \frac{\partial \mu}{\partial x} \frac{\partial}{\partial k_x}\right) f(x, k_x) =  0
\label{eq:boltzmann}
\end{equation}
where $f$ is the phase space distribution function. It is solved by the method of characteristics by assuming solutions of the form
$$
f(x, k_x) = g(\varepsilon(k_x) + \mu(x)).
$$
i.e. in the absence of collisions the distribution is transported along curves of constant energy. In order to get consistent solutions to the Boltzmann equation we are allowed to specify an arbitrary distribution that will be transported along the characteristics. The surplus electrons will come from the left reservoir at $x=0$ and travel to the right reservoir at $x=L$. To model the left reservoir we specify Fermi-Dirac statistics for right movers at chemical potential $\mu_L$:
$$
f(0, k) = n_F(\varepsilon(k_x) + \mu_L) \textrm{ for } v_x =\frac{1}{\hbar} \frac{\partial \varepsilon}{\partial k_x} > 0
$$
and vice-versa for the left movers coming from the right reservoir at chemical potential $\mu_R$:
$$
f(L, k) = n_F(\varepsilon(k_x) + \mu_R) \textrm{ for } v_x = \frac{1}{\hbar} \frac{\partial \varepsilon}{\partial k_x} < 0.
$$
Transporting the distribution on the right boundary along the characteristics to the left boundary gives the full distribution at $x = 0$:
\begin{align}
f(0, k) &= n_{F}(\varepsilon(k_x) + \mu_L) \Theta(v_x) + n_{F}(\varepsilon(k_x) + \mu_R) \Theta(-v_x) \nonumber \\
&= n_{F}(\varepsilon(k_x) + \mu_R + \Delta \mu) \Theta(v_x) + n_{F}(\varepsilon(k_x) + \mu_R) \Theta(-v_x) \nonumber \\
&= n_{F}(\varepsilon(k_x) + \mu_L) + \left( \frac{\partial n_F}{\partial \varepsilon} \Delta \mu + \frac{1}{2}\frac{\partial^2 n_F}{\partial \varepsilon^2}(\Delta \mu)^2  + \cdots \right)\Theta(v_x) \label{eq:boundary_condition}
\end{align}
This is the appropriate boundary condition for a perturbative expansion of Equation \ref{eq:boltzmann} in powers of $\Delta \mu$. Substituting a the constant electric field, group velocity, and expanding in powers of $\Delta \mu$, we have
\begin{equation*}
\left( \frac{\partial \varepsilon}{\partial k} \frac{\partial}{\partial x} - \frac{\Delta \mu}{L} \frac{\partial}{\partial k}\right) \left( \sum_{n=0}^{\infty} (\Delta \mu)^n f_n\right) =  0.
\end{equation*}
Grouping terms of order $1$ gives
\begin{align*}
\frac{\partial \varepsilon}{\partial k} \frac{\partial}{\partial x} f_{0} = 0
\end{align*}
Integrating and plugging in the boundary condition gives
\begin{align*}
    f_0 = n_{F}(\varepsilon(k_x) + \mu_L).
\end{align*}
At next order, we obtain
\begin{align*}
&\frac{\partial \varepsilon}{\partial k} \frac{\partial }{\partial x} f_1 - \frac{1}{L}\frac{\partial n_F}{\partial k} = 0.
\end{align*}
The second term is not small and cannot be neglected. Integrating this equation gives
\begin{align*}
& f_1 = \frac{x}{L} \frac{\partial n_F}{\partial \varepsilon} + C
\end{align*}
where $C$ is determined by the boundary condition at $x = 0$, resulting in
\begin{align*}
f_1 = \left( \frac{x}{L} + \Theta(v_x)\right) \frac{\partial n_F}{\partial \varepsilon}
\end{align*}
Finally at next order we have
\begin{align*}
&\frac{\partial \varepsilon}{\partial k} \frac{\partial }{\partial x} f_2 - \frac{1}{L} \frac{\partial}{\partial k} \left( \left( \frac{x}{L} + \Theta(v_x)\right) \frac{\partial n_F}{\partial \varepsilon} \right) = 0 \\
\Rightarrow &\frac{\partial \varepsilon}{\partial k} \frac{\partial }{\partial x} f_2 - \frac{1}{L} \left( \left( \frac{x}{L} +  \Theta(v_x)\right) \frac{\partial^2 n_F}{\partial \varepsilon^2} \frac{\partial \varepsilon}{\partial k}\right) - \frac{\partial n_F}{\partial \varepsilon} \delta(v_x) \frac{\partial v_x}{\partial k}= 0.
\end{align*}
We neglect the last term in the same short-wavelength approximation as above. Integrating and using the boundary condition we have
\begin{align*}
&f_2 = \left( \left( \frac{x}{L}\right)^2 + 2\frac{x}{L} \Theta(v_x)  + \Theta(v_x) \right) \frac{1}{2} \frac{\partial^2 n_F}{\partial \varepsilon^2}.
\end{align*}
The total surplus distribution is
\begin{align*}
\delta f &= \Delta \mu f_1 + \Delta \mu^2 f_2 \\
&=  \left( \frac{x}{L} + \Theta(v_x)\right) \frac{\partial n_F}{\partial \varepsilon} \Delta \mu + \left( \left( \frac{x}{L}\right)^2 + 2\frac{x}{L} \Theta(v_x)  + \Theta(v_x) \right) \frac{1}{2} \frac{\partial^2 n_F}{\partial \varepsilon^2} (\Delta \mu)^2
\end{align*}
To calculate the current we need to integrate the distribution over $x$:
\begin{align*}
L \int_0^1 \delta f d(x / L) &= L \left( \frac{1}{2} + \Theta(v_x)\right) \frac{\partial n_F}{\partial \varepsilon} \Delta \mu + L\left( \frac{1}{3} + \Theta(v_x)  + \Theta(v_x) \right) \frac{1}{2} \frac{\partial^2 n_F}{\partial \varepsilon^2} (\Delta \mu)^2 \\
&= L \left( \frac{1}{2} \frac{\partial n_F}{\partial \varepsilon} \Delta \mu + \frac{1}{6} \frac{\partial^2 n_F}{\partial \varepsilon^2} (\Delta \mu)^2\right) + L  \left(  \frac{\partial n_F}{\partial \varepsilon} \Delta \mu + \frac{\partial^2 n_F}{\partial \varepsilon^2} (\Delta \mu)^2 \right)\Theta(v_x)
\end{align*}
The first term is the shift in the distribution due to the local chemical potential and can be thought of as a bulk effect. The second is the surplus of right-movers coming from the left reservoir. 
The $y$-current is
\begin{align*}
I_y &= e \int_{0}^L dx \int_{BZ} \frac{d^2 k}{(2\pi)^2} \dot{y} \delta f \\
\end{align*}
where we note that $\dot{y}$ is time reversal odd. Therefore, the contribution from the first term is zero in time-reversal symmetric materials, as we would expect from a bulk effect. Finally, we find the current to be
\begin{align*}
I_y &= eL\frac{\Delta \mu}{2} \int_{BZ} \frac{d^2 k}{(2\pi)^2}\left(\frac{1}{\hbar}\frac{\partial \varepsilon}{\partial k_y} - \frac{\Delta \mu}{\hbar L}\Omega_z \right) \left( \frac{\partial n_F}{\partial \varepsilon}\Delta \mu + \frac{\partial^2 n_F}{\partial \varepsilon^2}(\Delta \mu)^2  + \cdots \right)\Theta( v_x)
\end{align*}
The transverse conductance is therefore
\begin{equation*}
G_{yx} = \frac{e}{\Delta \mu} I_y = \frac{e^2}{\hbar} \frac{L}{\Delta \mu} \int \frac{d^2 k}{(2\pi)^2} \left(\frac{\partial \varepsilon}{\partial k_y} - \frac{\Delta \mu}{L}\Omega_z \right) \left( \frac{\partial n_F}{\partial \varepsilon} \Delta \mu + \frac{\partial^2 n_F}{\partial \varepsilon^2}(\Delta \mu)^2  + \cdots \right)\Theta(k)
\end{equation*}
Organizing order by order in $\Delta \mu$ we get
\begin{align}
\begin{split}
G_{yx} = \frac{e^2}{\hbar} \int \frac{d^2 k}{(2\pi)^2} \Theta(v_x) &\bigg( L\frac{\partial n_F}{\partial \varepsilon} \frac{\partial \varepsilon}{\partial k_y} \\
&+ \left(L\frac{\partial^2 n_F}{\partial \varepsilon^2} \frac{\partial \varepsilon}{\partial k_y}  + \frac{\partial n_F}{\partial \varepsilon} \Omega_z \right) \Delta \mu \\
&+ \cdots \bigg)
\end{split}
\label{eq:full_response}
\end{align}
The angular dependence on $\theta$ can then be obtained by change of coordinates. We are interested in the linear term, which could be obtained by varying the potential and fitting the quadratic response. Then we need to subtract the other term at the same order. The error in the subtraction process and the fitting of the linear component of $G_{yx}$ would be accounted for in the estimation of the noise scale $\alpha$.

Now we need to address the fact that the terms we are not interested in are proportional to $L$, since they arise from terms independent of the electric field in the equations of motion. The maximum length for $L$ is set by the mean free path, since we need to be in the ballistic limit. The minimum length is set by the validity of the Boltzmann transport picture. This is estimated by calculating the thermal wavelength, which is
\begin{equation*}
\lambda \approx \frac{\hbar}{\sqrt{m^* k_B T}} = \frac{1}{\sqrt{m^* / m_e}}\frac{\hbar}{\sqrt{m_e  k_B T}} = \frac{1}{\sqrt{m^* / m_e}} \times 20 \textrm{ nm}
\end{equation*}
For WSe$_2$, $m^* / m_e = 0.35$ in the upper conduction band ~\cite{Le2015WSe2SOC} so that the minimum device size is roughly 35 nm. To minimize the effects of non-Berry curvature terms this should be the characteristic size of the device. 

\section{Discretization}
\noindent
\textbf{Discrete Integral Operator}. First we describe how to discretize the integral operator $A$. We choose a discretization of the relevant parts of the Brioullin zone (BZ) labelled by $\mathbf{k}_j$ with $j = 1, \cdots, m$ and a set of angles and chemical potentials $(\theta, \mu)_{i}$ with $i=1,\cdots, n$ labeling the measurements. Along with these we have a vector of integration weights $w(\mathbf{k}_j)$. We want a linear operator that acts on $\Omega$ only. When we do one integral for $G_H$ we have
\begin{equation*}
\frac{e^2 \Delta \mu}{h \pi} \int_{BZ} d^2k \left( -\frac{\partial n_F}{\partial \varepsilon}(\varepsilon(\mathbf{k}) - \mu)\right) \Theta(-\mathbf{v}(\mathbf{k}) \cdot \mathbf{E}(\theta)) \Omega_z(\mathbf{k}) \rightarrow \sum_{j} w(\mathbf{k}_j) V((\mu, \theta)_i, \mathbf{k}_j) \Omega(\mathbf{k}_j) 
\end{equation*}
where $V((\mu, \theta)_i, \mathbf{k}_j)$ has entries
\begin{equation*}
V((\mu, \theta)_i, \mathbf{k}_j) = \frac{e^2 \Delta \mu}{h \pi} \left(-\frac{\partial n_F}{\partial \varepsilon}(\varepsilon(\mathbf{k}_j) - \mu)\right)\Theta(-\mathbf{v}({\mathbf{k}_j}) \cdot \mathbf{E}(\theta)).
\end{equation*}
The discrete measurement operator has entries given by
\begin{equation*}
[A]_{ij} = \sum_{j} w(\mathbf{k}_j) V((\mu, \theta)_i, \mathbf{k}_j) .
\end{equation*}
\textbf{Symmetry}. We can exploit symmetry by taking measurements at symmetry equivalent angles to average over the noise in each measurement. Let $\mathbf{k}_j'$ be the set of points corresponding to our mesh in an irreducible wedge $\Gamma$ of the Brioullin zone. What this means is that there is a set of maps $R_a$ that unfold the wedge, i.e. take any given $k$ point in $\Gamma$ and assign them to a point in the full Brioullin zone. The structure of $V$ is then modified as
\begin{equation*}
V((\mu, \theta)_i, \mathbf{k}_j') = \frac{e^2 \Delta \mu}{h \pi}\sum_{a} \left(-\frac{\partial n_F}{\partial \varepsilon}(\varepsilon(R_a \mathbf{k}_j') - \mu)\right)\Theta(- \mathbf{v}(R_a \mathbf{k}_j') \cdot \mathbf{E}(\theta)) S({R_a})
\end{equation*}
\begin{equation*}
\tilde{w}(\mathbf{k}_j') = \sum_{a} w(R_a\mathbf{k}'_j)
\end{equation*}
so that
\begin{equation*}
[A]_{ij} = \sum_{j} \tilde{w}(\mathbf{k}'_j) V((\mu, \theta)_i, \mathbf{k}'_j).
\end{equation*}
\textbf{Regularizer}. Lastly, for the regularizer we formulate a finite difference Laplacian which obeys the exact conservation law with respect to the integration weights. In both materials choose a polar coordinate system around the $K$ point. After including all the symmetries into our discrete integral operator, we are left with an irreducible wedge, which we parameterize in the polar coordinate system. On the wedge we have a diagonal matrix $V = \textrm{diag}(w(\mathbf{k}'_j))$ with the integration weights for the given points. Here we use the integration weights corresponding to the area elements on the wedge, not the weights $\tilde w$ corresponding to the symmetry unfolded Brioullin zone. The corresponding Laplacian is
\begin{equation*}
L =  -V^{-1} D_{r}^T V D_{r} - V^{-1} D_{\theta}^T V D_{\theta} 
\end{equation*}
where $D_r$ and $D_\theta$ are the finite-difference matrices for the $r$- and $\theta$- derivatives obeying Neumann boundary conditions on all boundaries. We chose first order forward differences for $D_r$ and $D_\theta$.

\section{Bayesian Formulation}
Now that we have the data generating operators we can make the inverse model. A weighted least-squares procedure is computationally efficient and conceptually simple. We assume the measurements $G^o$ are given as
\begin{equation}
G^o = A \Omega + \epsilon
\end{equation}
where $\epsilon \sim \mathcal{N}(0, (\alpha W)^{-1})$ is a zero mean Gaussian random variable with covariance $(\alpha W)^{-1}$, where $\alpha$ is a scaling parameter for the noise model that will be estimated from the data. 
\\\\
\noindent
\textbf{Likelihood}. We have therefore that the likelihood over $G^o$ is given by
\begin{equation*}
P(G^o | \Omega, \alpha) = \sqrt{\frac{\det \alpha W}{(2\pi)^n }}\exp\left(-\frac{\alpha}{2} (G^o - A\Omega)^T W(G^o - A\Omega)\right)
\end{equation*}
where $n$ is the length of the vector $G^o$, i.e. the number of measurements.
\\\\
\noindent
\textbf{Prior}. We assign the following factored form for the prior distribution
\begin{align*}
P(\Omega, \alpha, \gamma) &= P(\Omega | \gamma) P(\gamma) P(\alpha) \\
\end{align*}
where we assign a gaussian distribution for the probability on $\Omega$ conditioned on $\gamma$:
\begin{align*}
P(\Omega | \gamma) &= \sqrt{\frac{\det \gamma L^T L}{(2\pi)^m}} \exp\left(-\frac{1}{2} \gamma (L \Omega)^T(L\Omega) \right)
\end{align*}
where $\gamma$ sets the width of the prior distribution (strength of the regularizer), and $\alpha$ and $\gamma$ are assigned a flat prior. Here $m$ is the number of $k$-points on the reduced mesh.
\\\\
\noindent
\textbf{Posterior}. The posterior distribution is given by Bayes's rule
\begin{align*}
P(\Omega, \alpha, \gamma | G^o) &= \frac{P(G^o | \Omega, \alpha) P(\Omega, \alpha, \gamma)}{P(G^o)} \\
&= \frac{1}{Z} \sqrt{\frac{\det \alpha W}{(2\pi)^n}}\exp\left(-\frac{\alpha}{2} (G^o - A\Omega)^T W (G^o - A\Omega)\right) \\
&\times \sqrt{\frac{\det \gamma L^T L}{(2\pi)^m}}\exp\left(-\frac{\gamma}{2} (L \Omega)^T(L\Omega) \right)
\end{align*}
where $Z$ is a normalizing constant. We proceed in two steps by factoring the posterior
\begin{align*}
P(\Omega, \alpha, \gamma | G^o) = P(\Omega | \alpha, \gamma, G^o) P(\alpha, \gamma | G^o).
\end{align*}
Now we calculate the expected value of $\Omega$ conditioned on $\alpha$ and $\gamma$, which is straightforward because it is gaussian:
\begin{align*}
\bar \Omega(\alpha, \gamma) &\equiv \mathbb{E}[P(\Omega | \alpha, \gamma, G^o)]\\
&= \underset{\Omega}{\textrm{argmax}} \left(- \frac{\alpha}{2} (G^o - A\Omega)^T W (G^o - A\Omega) - \frac{\gamma}{2}(L \Omega)^T(L\Omega) \right).
\end{align*}
The saddle-point equations for this maximization problem are
\begin{align*}
& \nabla _{\Omega}P(\Omega| \alpha, \gamma, G^o) |_{\bar \Omega}  = 0 \\
&\Rightarrow  - \alpha A^T W G^o + \alpha A^T W A \bar \Omega  + \gamma L^T L \bar \Omega = 0 \\
&\Rightarrow \bar \Omega(\alpha, \gamma) = ( A^T W A + (\gamma / \alpha) L^T L)^{-1} A^T W G^o.
\end{align*}
As far as the mean Berry curvature goes our statistical model is characterized by the ratio $\gamma / \alpha$. As discussed in the next section, typically the full operator with the inclusion of the prior is invertible. In a following section we calculate $P(\alpha, \gamma | G^o)$ to set the value of the hyperparameters. 

\subsection{Invertibility and Spectra of the Measurement Operator}

\begin{figure}
    \centering
    \includegraphics[width=0.9\linewidth]{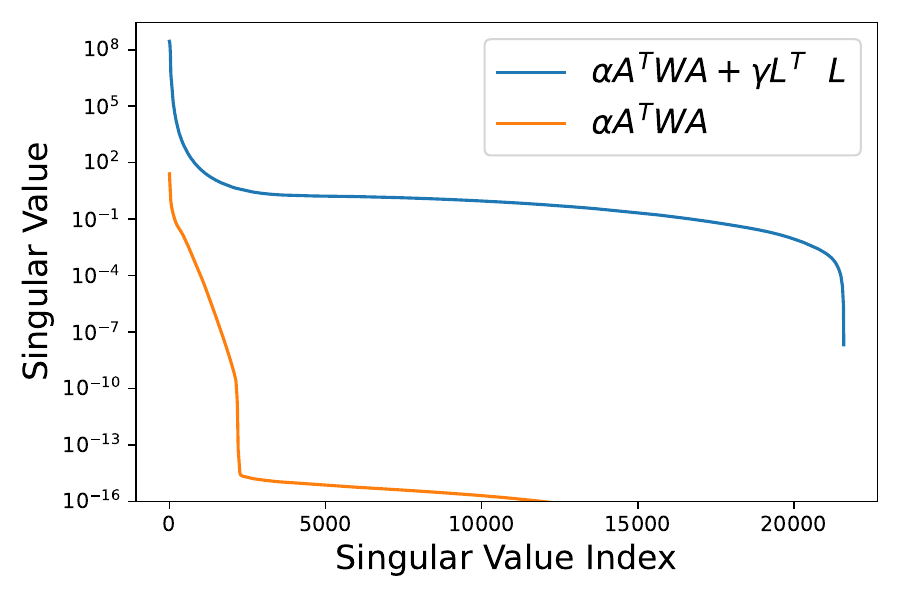}
    \caption{Singular values of the measurement operator with and without regularization for the inversion performed in Figure \ref{fig:abc}. The smoothness prior lifts all singular values above the floating point precision cutoff. }
    \label{fig:svd}
\end{figure}

Because the operator $A^T W A$ is singular (see Figure \ref{fig:svd} for the singular values), it is not obvious that the problem is solvable at all. However, the smoothness prior renders the total operator invertible. Hence, we conclude that the ballistic Hall effect contains enough information to determine the Berry curvature, since we know it is a smooth function. 
We can give a rough linear algebraic argument. We took as a regularizer a Laplacian on a simply connected domain (the irreducible wedge) with Neumann boundary conditions on all boundaries. Hence it has one null mode which is the constant field. But the constant field is not in the (right) null space of the measurement operator $A$, so long as the Fermi surface is anisotropic. Hence the two positive-semidefinite operators $A^T W A$ and $L^T L$ have no shared null eigenspaces and their convex combination is strictly positive definite and invertible. 

\subsection{Automatic Hyperparameter Tuning}

In order to automatically set the strength of the prior, we can calculate the MAP estimate for $\alpha$ and $\gamma$. To do so, we want to calculate the marginal posterior $P(\gamma, \alpha | G^o)$. This is given by Bayes's rule:
\begin{equation*}
P(\gamma, \alpha | G^o) \propto P(G^o | \gamma, \alpha) P(\gamma, \alpha)
\end{equation*}
The marginal log likelihood is obtained by integrating out $\Omega$:
\begin{align*}
P(G^o |\gamma, \alpha) &= \int P(G^o, \Omega | \gamma, \alpha) d\Omega \\
&= \int P(G^o | \Omega, \gamma, \alpha)P(\Omega | \gamma, \alpha) d\Omega \\
&= \sqrt{\frac{\det \alpha W }{(2\pi)^n}} \sqrt{\frac{\det \gamma L^T L}{(2\pi)^m}}\int \exp\left(-\frac{1}{2} \left[ (G^o - A\Omega)^T \alpha W (G^o - A\Omega) + \gamma (L \Omega)^T(L\Omega) \right] \right) d\Omega\\
&= \sqrt{\frac{\det \alpha W}{(2\pi)^n}} \sqrt{\frac{\det \gamma L^T L}{(2\pi)^m}} \sqrt{\frac{(2\pi)^m}{\det \Lambda} }\exp\left(-\frac{1}{2} (G^o)^T[\alpha W - \alpha W^T A \Lambda^{-1}A^T \alpha W ]G^o\right)
\end{align*}
with
\begin{equation*}
\Lambda =  \alpha A^T W A + \gamma L^T L
\end{equation*}
Since we have flat priors on $\alpha$ and $\gamma$, the marginal log posterior is 
\begin{align*}
\log P(\alpha, \gamma | G^o) &= \frac{1}{2} \log \det \alpha W + \frac{1}{2}\log\det \left(\gamma L^T L \right) - \frac{1}{2}\log \det\Lambda \\
&- \frac{1}{2}(G^o)^T \left( \alpha W - \alpha^2 W^T A \Lambda^{-1}A^T W \right) G^o.
\end{align*}
We can drop constant factors to get a more computationally efficient version:
\begin{align*}
\log P(\gamma, \alpha | G^o) &= \frac{n}{2} \log\alpha + \frac{m}{2} \log\gamma - \frac{1}{2}\log \det\Lambda \\
&- \alpha \left(\frac{1}{2}(G^o)^T W G^o\right) - \frac{1}{2} \bar\Omega \Lambda \bar{\Omega}
\end{align*}
which is an objective function that we can maximize. We arrive finally at the full expression for the reconstruction:
\begin{align*}
\hat{\Omega} = \bar{\Omega}(\hat{\alpha}, \hat{\gamma})= ( A^T W A + (\hat{\gamma} / \hat{\alpha}) L^T L)^{-1} A^T W G^o. \\
\hat{\alpha}, \hat{\gamma} = \textrm{argmax}_{\alpha, \gamma} \log P(\gamma, \alpha | G^o)
\end{align*}
This is the MAP estimate over the whole posterior. 

\begin{figure}
    \centering
    \includegraphics[width=1.0\linewidth]{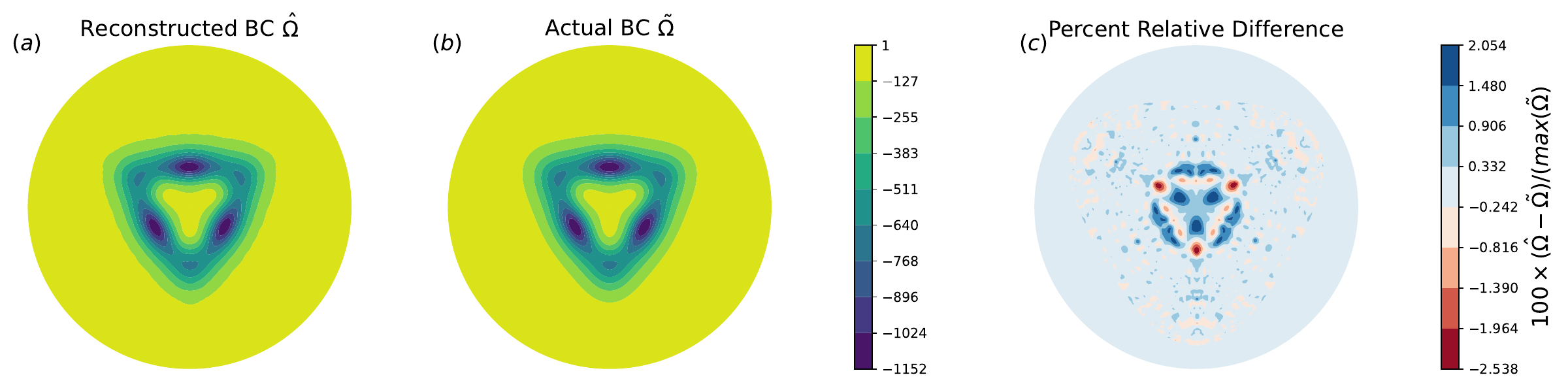}
    \caption{Relative difference of the reconstructed Berry curvature and the actual Berry curvature with respect to the maximum absolute Berry curvature in the domain.}
    \label{fig:relative_difference_abc}
\end{figure}

\begin{figure}
    \centering
    \includegraphics[width=0.8\linewidth]{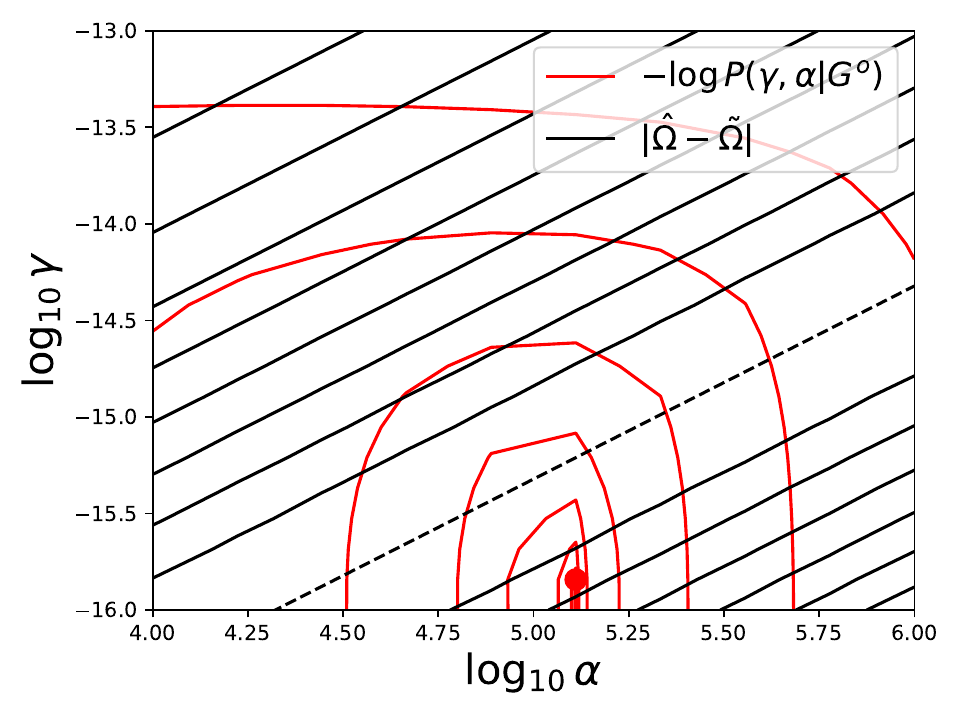}
    \caption{In red, a plot of the negative of the log-likelihood function (Equation \ref{eq:marginal_log_likelihood}) used for the reconstruction in Figure \ref{fig:abc}. The contour values are chosen on a log scale for visual clarity. In black we plot the distance to the true solution $\tilde{\Omega}$ in the Euclidean norm (not weighted by the polar coordinate area elements). Since the prediction only depends on $\gamma / \alpha$ this function is constant along contours of constant $\gamma / \alpha$. The minimum of $|\hat{\Omega} - \tilde \Omega|$ is plotted with a dashed line, while the red dot indicates the minimum of the negative marginal log posterior. We note empirically that the curvature of $|\tilde \Omega - \hat \Omega|$ becomes steeper when the noise level is higher, so that disagreements between the two minima are penalized more heavily. }
    \label{fig:log_likelihood}
\end{figure}

The marginal log posterior is plotted in Figure $\ref{fig:log_likelihood}$ for the reconstruction of $ABC$-stacked graphene (Figure \ref{fig:abc}) with a realization of the noise drawn from $\mathcal{N}(0, 3 \times 10^{-3} \frac{1}{n}\sum|G^o|I)$, i.e. all measurements were corrupted with gaussian noise with standard deviation proportional to the average of the absolute value of the true signal. The reconstruction is reproduced in figure \ref{fig:relative_difference_abc}, along with the relative difference with respect to the maximum Berry curvature. $W$ was set to be the identity times the average of the absolute value of the true signal, so that the task was to estimate $\alpha = (3 \times 10^{-3})^{-2}$. The grid search that generated the contour plot in Figure \ref{fig:log_likelihood} shows that the maximum of the log posterior is appropriately placed at $\hat{\alpha} = 10^5$, which is close to the ``true" value, ignoring contributions from residual misfit. The value of $\hat{\gamma}$ is in turn also close to the optimal value, leading to a good reconstruction. 

This marginalization procedure can be interpreted as a form of renormalization. The prior sets the asymptotic high-frequency behavior, and $\gamma$ acts as the scale parameter. In order to set $\gamma$, we integrate out the field $\Omega$, which is possible exactly since we have a quadratic model. Finding the maximum leads to a stationarity condition for the optimal $\alpha$ which is equivalent of the renormalization group equation. The new $\alpha$ differing from $1$ is the renormalized coupling constant. 

Finally, we can use the marginal posterior for $\Omega$ to generate samples and assess the error in our reconstruction. The marginal posterior over the Berry curvature $\Omega$ is
\begin{equation}
    P(\Omega|\gamma, \alpha, G^o) = \mathcal{N}(\bar{\Omega}(\hat{\alpha}, \hat{\gamma}), (\hat{\alpha} A^T W A + \hat{\gamma} L^T L)^{-1})
\end{equation}
which we can sample efficiently by computing the Cholesky factors of the covariance.

\begin{figure}
    \centering
    \includegraphics[width=\linewidth]{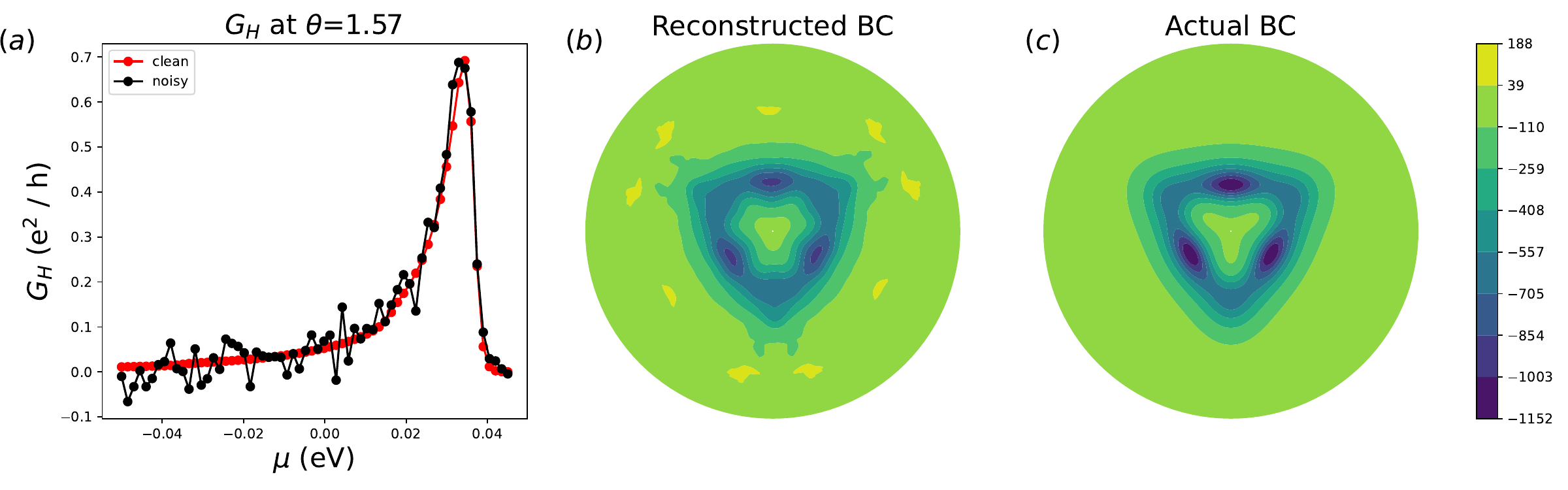}
    \caption{Reconstruction for the noise standard deviation 1/2 of the average clean signal, shown in figure (a) for a chemical potential sweep at the angle that gives the maximum response. In (b) the Berry curvature is qualitatively similar to the true Berry curvature (c) from the tight-binding model.}
    \label{fig:noisy_reconstruction}
\end{figure}

\subsection{Noise and Measurement Resolution Dependence}
\begin{figure}
    \centering
    \includegraphics[width=0.8\linewidth]{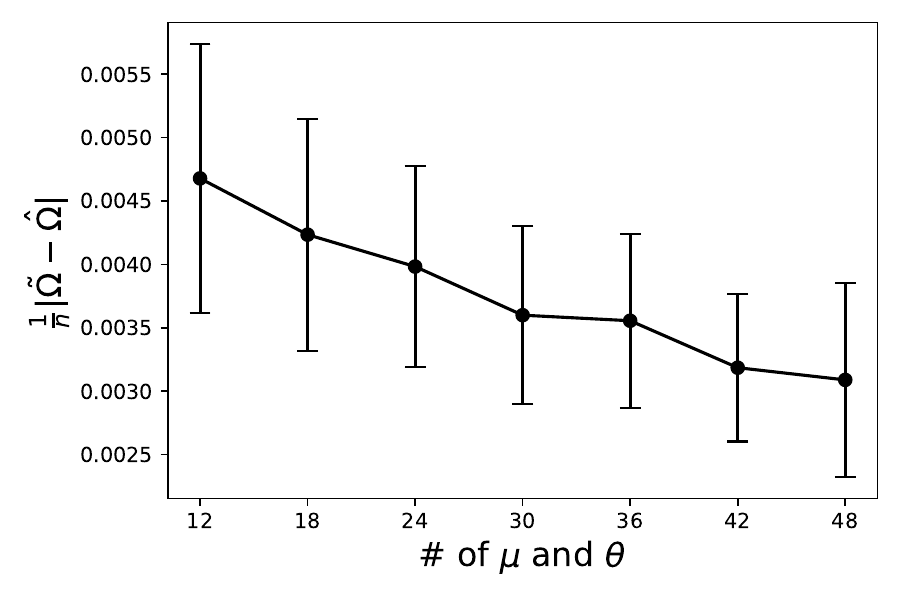}
    \caption{Mean error ($n$ is the number of irreducible $k$-points) for differing numbers of measurements of $\mu$ and $\theta$. 20 trials were averaged to get the mean and standard deviation, which give the error bars in the figure. We remark here that as we have sketched the measurement setup, the number of angle resolved measurements should always be a multiple of four in order to have measurement leads orthogonal to the injecting leads, although we did not obey this rule in this plot.}
    \label{fig:error_scaling}

\end{figure}

As a final note, we show that the reconstruction can be improved by including more measurements. For this numerical experiment we use a coarse mesh for the Berry curvature due to computational cost. In this example we use the model of abc-graphene, with the same parameters (except for the $k$-mesh) as above. We consider measurement setups with varying resolution in $\mu$ and $\theta$, plotted in Figure \ref{fig:error_scaling}. We perform inversions for 20 realizations of the noise, maximizing the log posterior in each to calculate $\hat{\alpha}$ and $\hat{\gamma}$. Increasing the number of measurements improves the error in the Berry curvature relative to the true solution. 

\end{document}